\documentclass{IEEEtran}

\ifCLASSINFOpdf
\else
\fi

\usepackage{amsthm}
\usepackage{graphicx}
\usepackage{amsmath}
\usepackage{amssymb}
\usepackage{mathenv}
\usepackage{subfigure}
\usepackage{mathrsfs}
\usepackage{graphics}
\usepackage{epsfig}
\usepackage{epstopdf}
\usepackage{multirow}
\usepackage{cite}
\usepackage{enumitem}

\theoremstyle{plain}

\begin{document}
\bstctlcite{BSTcontrol}
\title{Distributed and Joint Optimization of Precoding and Power for User-Centric Cell-Free Massive MIMO}
\author{\IEEEauthorblockN{Hongkang Yu, Xinquan Ye, Yijian Chen}}
\maketitle

\begin{abstract}
	In the cell-free massive multiple-input multiple-output (CF mMIMO) system, the centralized transmission scheme is widely adopted to manage the inter-user interference. Unfortunately, its implementation is limited by the extensive signaling overhead between the central process unit (CPU) and the access points (APs). To solve this problem, we propose a distributed downlink transmission scheme in this letter. First, the null space-based precoding is used to cancel the interference to partial users, where only a portion of channel state information (CSI) needs to be shared among the AP cluster.	Based on this, the dual decomposition method is adopted to jointly optimize the precoder and power control, where the calculation can be performed independently by each AP cluster with closed-form expression. With very few iterations, our distributed scheme achieves the same performance as the centralized one. Moreover, it significantly reduces the information exchange to the CPU.



\end{abstract}


\section{Introduction}
\IEEEPARstart{C}{ell-free} massive multiple-input multiple-output (CF mMIMO) is a promising technology for the next-generation mobile communication network \cite{ref1}. Compared to the traditional cellular system, the CF system relies on a large number of distributed access points (APs), which are connected to the central process unit (CPU) and serve all users in a cooperative approach. To further reduce signaling overhead and complexity, a user-centric CF concept was introduced in \cite{ref3}, where each user is only served by neighboring APs. As a result, this scheme can deliver uniformly good service and has scalability even in a large-scale network \cite{ref4}.

When considering the downlink transmission of the CF system, the distributed scheme, e.g., conjugate beamforming (CB), was first proposed \cite{ref5}. It requires only local channel state information (CSI) to design the precoder but cannot suppress interference effectively. As a comparison, the centralized scheme, e.g., zero-forcing (ZF), usually has a better performance \cite{ref9}. However, its implementation is limited by the extensive information exchange between the CPU and the APs, which includes both CSI and data payload \cite{ref1, ref15}.

To solve this problem, a fully distributed local partial ZF (PZF) scheme was proposed in \cite{ref7}. However, it requires a large number of antennas per AP, usually more than the number of users.
In \cite{ref8}, an over-the-air CSI exchange mechanism was introduced to reduce the signaling overhead, which requires extra time-frequency resources and therefore reduces the spectral efficiency (SE). In \cite{ref10}, the authors proposed to divide all APs into multiple disjoint clusters, and the precoder is calculated among each cluster. As a result, the users at the cluster edge may suffer severe interference.

Power control is another key issue when designing the transmission scheme. Although this topic has been well-discussed, the following problems still exist. First, most studies optimize the power with a general-purpose convex optimization solver \cite{ref7}. To reduce the complexity, algorithms with closed-form equations were derived in \cite{ref11}, which still involves multiple iterations. Moreover, the above schemes all perform power control after determining the precoding vector, and this decoupling method may result in a performance loss.


In this letter, we focus on the distributed and joint design of the precoder and power control for the user-centric CF system. First, a reduced CSI exchange mechanism is proposed, where CSI is only shared by neighboring APs. Based on this, we adopt the null space-based precoding to eliminate the interference to a portion of users. Specifically, we use dual decomposition method to jointly optimize the transmission scheme, so that each AP cluster can   independently calculate the precoding vector with closed-form expression. Simulation results show that the proposed scheme outperforms the widely used decoupling method with very few iterations. Meanwhile, this scheme enjoys a lower signaling overhead and complexity.



\section{System Model}
This letter considers a user-centric CF network with $L$ APs and $K$ single-antenna users. Each AP has $N$ antennas and is connected to the CPU via the fronthaul. We denote the indices of the serving APs for the user $k$ as ${\mathcal{M}_k} \subset \{ {1, \ldots ,L} \}$ and assume that all ${\mathcal{M}_k}$ have the same size without loss of generality. Conventionally, the CPU can determine ${\mathcal{M}_k}$ according to the pathloss between the APs and the users.

We assume that the system operates in a time division duplexing (TDD) mode, and block-fading channel model is adopted, where each coherence interval is divided into two phases: ${\tau _{\text{p}}}$ channel uses are dedicated for the uplink pilot, and the remaining ${\tau _{\text{d}}}$ channel uses for the downlink data. During each coherence interval, the channel vector ${{\mathbf{h}}_{k,l}}\sim\mathcal{C}\mathcal{N}\left( {0,{{\mathbf{R}}_{k,l}}} \right)$ between the AP $l$ and the user $k$ stays constant, where ${{\mathbf{R}}_{k,l}}$ is the spatial correlation matrix, and ${\beta _{k,l}} = {\text{Tr}}\left( {{{\mathbf{R}}_{k,l}}} \right)/N$ denotes the pathloss. By exploiting channel reciprocity, AP can obtain downlink CSI by performing channel estimation from the uplink pilot. Since this letter focuses on the joint design of the precoder and power control, we assume that the perfect CSI is available, and this can be achieved under high signal-to-noise ratio (SNR) condition and a reasonable pilot allocation strategy \cite{ref12}. Moreover, the proposed scheme only requires the AP to obtain the CSI of neighboring users, which is more practical and will be described in detail in the next section.

Let ${s_k}$ denote the symbol sent to the user $k$, which satisfies $\mathbb{E}\{ {{{\left| {{s_k}} \right|}^2}} \} = 1$. The transmitted signal at the AP $l$ can be expressed as
\begin{equation}
	{{\mathbf{x}}_l} = \sum\limits_{k \in {\mathcal{D}_l}} {\sqrt {{\rho _{k,l}}} {{\mathbf{w}}_{k,l}}{s_k}},
\end{equation}
where ${\mathcal{D}_l}$ represents the set of users served by AP $l$, ${{\mathbf{w}}_{k,l}}$ denotes the corresponding precoding vector that has unit power, i.e., $\left\| {{{\mathbf{w}}_{k,l}}} \right\|_2^2 = 1$, and ${\rho _{k,l}}$ can be seen as the power allocated for the user $k$. Assuming that perfect synchronization can be realized in the system, the received signal at the user $k$ can be modeled as
\begin{eqnarray}
	\begin{aligned}
		{y_k} & =\!  \sum\limits_l {{\mathbf{h}}_{k,l}^{\text{H}}{{\mathbf{x}}_l}} +{n_k}                                                                                                                                                                                                                     \\
		      & =\!\sum\limits_{l \in {\mathcal{M}_k}}\! {\sqrt {{\rho _{k,l}}}{\mathbf{h}}_{k,l}^{\text{H}}{{\mathbf{w}}_{k,l}}{s_k}}\!+\!\sum\limits_{k' \ne k}\! {\sum\limits_{l \in {\mathcal{M}_{k'}}} \!{\sqrt {{\rho _{k',l}}}{\mathbf{h}}_{k,l}^{\text{H}}{{\mathbf{w}}_{k',l}}{s_{k'}}} }\!+\!{n_k}, \\
	\end{aligned}
\end{eqnarray}
where ${n_k}\sim\mathcal{C}\mathcal{N}\left( {0,1} \right)$ represents the normalized noise. Based on this model, the signal-to-interference-and-noise ratio (SINR) is given by
\begin{equation}
	{\text{SINR}}_k = \frac{{{{\left| {\sum\limits_{l \in {\mathcal{M}_k}} {\sqrt {{\rho _{k,l}}} {\mathbf{h}}_{k,l}^{\text{H}}{{\mathbf{w}}_{k,l}}} } \right|}^2}}}{{\sum\limits_{k' \ne k} {{{\left| {\sum\limits_{l \in {\mathcal{M}_{k'}}} {\sqrt {{\rho _{k',l}}} {\mathbf{h}}_{k,l}^{\text{H}}{{\mathbf{w}}_{k',l}}} } \right|}^2}}  + 1}},
\end{equation}
and the sum-SE maximization problem can be formulated as
\begin{eqnarray}\label{original problem}
	\begin{aligned}
		\mathop {\max }\limits_{{{\mathbf{w}}_{k,l}},{\rho _{k,l}}}\;\;\  & \sum\limits_k {{{\log }_2}\left( {1 + {\text{SIN}}{{\text{R}}_k}} \right)}                      \\
		{\text{s.t.}}\;\;\                                                & \left\| {{{\mathbf{w}}_{k,l}}} \right\|_2^2 = 1,\forall l\;{\text{and}}\;k \in {\mathcal{D}_l}, \\
		                                                                  & \sum\limits_{k \in {\mathcal{D}_l}} {{{\rho _{k,l}}} }  \leqslant {\rho _{\max }},\forall l,    \\
	\end{aligned}
\end{eqnarray}
where ${\rho _{\max }}$ denotes the normalized maximum transmit power of each AP. Due to the highly non-convexity of the objective function, it is non-tractable to obtain the optimal solution, and the sequential optimization of ${{\mathbf{w}}_{k,l}}$ and ${\rho _{k,l}}$ may result in a performance loss. More importantly, the centralized scheme requires extensive information exchange between the CPU and the APs. Therefore, it is necessary to study the distributed and joint design of the precoder and power control.

\section{Transmission Scheme Design}
In this section, we first propose a reduced CSI exchange mechanism and the corresponding PZF strategy. Next, the transmission scheme is designed via both centralized and distributed approach.
\subsection{CSI Exchange Mechanism and PZF Precoding}\label{CSI exchange}
In a canonical CF system, all CSI is required to serve all users. However, when considering the user-centric CF system, AP clusters usually do not need to serve users that are far apart, and the corresponding CSI is no longer necessary. Motivated by this point, we propose a novel CSI exchange mechanism by defining the \emph{CSI sharing set} ${\mathcal{C}_k} \subseteq \{ 1, \ldots ,K\}$ for each user. When designing the precoder for the user $k$, only $\left\{ {{{\mathbf{h}}_{k,l}}:k \in {\mathcal{C}_k},l \in {\mathcal{M}_k}} \right\}$ is required. We can limit the size of ${\mathcal{C}_k}$ to balance the system performance and the overhead. Apart from the user $k$, the remaining $\left| {{\mathcal{C}_k} - 1} \right|$ users in ${\mathcal{C}_k}$ can be selected according to the average pathloss to APs in ${\mathcal{M}_k}$. As a result, only a small-scale CSI exchange is required between the APs, and the proposed distributed scheme even avoids sending CSI to the CPU, which will be introduced in section \ref{dis_scheme}.


\begin{figure}
	\centering
	\includegraphics[width=2in]{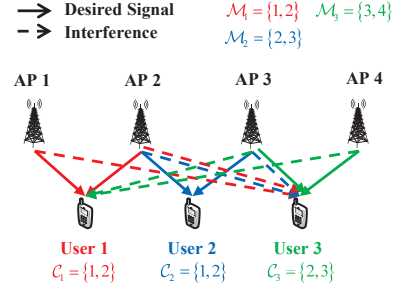}
	\caption{An illustration of the proposed transmission scheme.}
	\label{illustration}
\end{figure}

Based on the proposed CSI exchange mechanism, PZF precoding is adopted for the downlink transmission. As illustrated in Fig. \ref{illustration}, with the available CSI, the APs in ${\mathcal{M}_k}$ are designed to eliminate the interference to users in ${\mathcal{C}_k}\backslash k$. Although the interference to other users still remains, it has less important impact on the performance due to more severe pathloss. Next, we propose a null space-based transmission scheme, which realizes the joint optimization of the precoder and power control. Specifically, define

\begin{equation}
	{{\mathbf{\tilde H}}_k} = \left[ {\begin{array}{*{20}{c}}
					{{{\mathbf{h}}_{{i_1},{j_1}}}}                                  & \cdots & {{{\mathbf{h}}_{{i_{\left| {{\mathcal{C}_k}} \right| - 1}},{j_1}}}}                                  \\
					\vdots                                                          & \ddots & \vdots                                                                                               \\
					{{{\mathbf{h}}_{{i_1},{j_{\left| {{\mathcal{M}_k}} \right|}}}}} & \cdots & {{{\mathbf{h}}_{{i_{\left| {{\mathcal{C}_k}} \right| - 1}},{j_{\left| {{\mathcal{M}_k}} \right|}}}}}
				\end{array}} \right] \in {\mathbb{C}^{N\left| {{\mathcal{M}_k}} \right| \times \left( {\left| {{\mathcal{C}_k}} \right| - 1} \right)}}
\end{equation}
as the aggregated channel matrix for available CSI at ${\mathcal{M}_k}$, where ${i_n}$ and ${j_m}$ denotes the $n$-th element in ${\mathcal{C}_k}\backslash k$ and the $m$-th element in ${\mathcal{M}_k}$, respectively. We require system parameters to satisfy $N\left| {{\mathcal{M}_k}} \right| > \left| {{\mathcal{C}_k}} \right| - 1$ and perform singular value decomposition (SVD) on ${\mathbf{\tilde H}}_k^{\text{H}}$. Then, the null space can be obtained as
\begin{equation}
	{{\mathbf{N}}_k} \triangleq {\left[ {{\mathbf{N}}_{k,{j_1}}^{\text{T}}, \ldots ,{\mathbf{N}}_{k,{j_{\left| {{\mathcal{M}_k}} \right|}}}^{\text{T}}} \right]^{\text{T}}} \in {\mathbb{C}^{N\left| {{\mathcal{M}_k}} \right| \times \left( {N\left| {{\mathcal{M}_k}} \right| - \left| {{\mathcal{C}_k}} \right| + 1} \right)}},
\end{equation}
and the PZF precoder can be selected as a linear combination of ${{\mathbf{N}}_k}$ with coefficient ${{\mathbf{c}}_k}$, i.e.,
\begin{equation}\label{joint opt}
	{{\mathbf{w}}_k} \triangleq {\left[ {\sqrt {{\rho _{k,{j_1}}}} {\mathbf{w}}_{k,{j_1}}^{\text{T}}, \ldots ,\sqrt {{\rho _{k,{j_{\left| {{\mathcal{M}_k}} \right|}}}}} {\mathbf{w}}_{k,{j_{\left| {{\mathcal{M}_k}} \right|}}}^{\text{T}}} \right]^{\text{T}}} = {{\mathbf{N}}_k}{{\mathbf{c}}_k},
\end{equation}
which satisfies ${{\mathbf{\tilde H}}_k^\mathrm{H}}{{\mathbf{w}}_k} = {\mathbf{0}}$. Finally, the problem \eqref{original problem} can be transformed as
\begin{eqnarray}\label{approx problem}
	\begin{aligned}
		\mathop {\max }\limits_{{{\mathbf{c}}_k}} \;\; & \sum\limits_{k} {{{\log }_2}\left( {1 + {{\left| {{\mathbf{h}}_k^{\text{H}}{{\mathbf{N}}_k}{{\mathbf{c}}_k}} \right|}^2}} \right)}                                                         \\
		{\text{s}}{\text{.t}}{\text{.}}           \;\; & \sum\limits_{k \in {\mathcal{D}_l}} {{{\left\| {{{\mathbf{N}}_{k,l}}{{\mathbf{c}}_k}} \right\|}^2}}  \leqslant {\rho _{\max }},\;\;{\text{for}}\;\;l:\left| {{\mathcal{D}_l}} \right| > 0, \\
	\end{aligned}
\end{eqnarray}
where the interference that PZF does not eliminate is ignored, and the objective function is an approximation of the actual SE. According to \eqref{joint opt}, the joint optimization of the transmission scheme can be obtained by finding the optimal coefficient ${\mathbf{c}}_k^*$, which will be discussed in the next two subsections.

\subsection{Centralized Transmission Scheme}
This subsection briefly introduces the centralized transmission scheme, which is primarily used as a benchmark. In this scheme, the CPU is required to collect all CSI and to find ${\mathbf{c}}_k^*$ via semidefinite programming (SDP) method.

Specifically, by introducing semidefinite matrix ${{\mathbf{C}}_k} = {{\mathbf{c}}_k}{\mathbf{c}}_k^{\text{H}}$, problem \eqref{approx problem} can be reformulated as
\begin{eqnarray}\label{sdp problem}
	\begin{aligned}
		\mathop {\max }\limits_{{{\mathbf{C}}_k}} \;\; & \sum\limits_{k} {{{\log }_2}\left( {1 + {\text{Tr}}\left( {{\mathbf{N}}_k^{\text{H}}{{\mathbf{h}}_k}{\mathbf{h}}_k^{\text{H}}{{\mathbf{N}}_k}{{\mathbf{C}}_k}} \right)} \right)}                                           \\
		{\text{s}}{\text{.t}}{\text{.}}          \;\;  & \sum\limits_{k \in {\mathcal{D}_l}} {{\text{Tr}}\left( {{\mathbf{N}}_{k,l}^{\text{H}}{{\mathbf{N}}_{k,l}}{{\mathbf{C}}_k}} \right)}  \leqslant {\rho _{\max }},\;\;{\text{for}}\;\;l:\left| {{\mathcal{D}_l}} \right| > 0, \\
	\end{aligned}
\end{eqnarray}
which is a standard convex optimization problem and can be optimally solved via tools such as CVX. It should be noted that we drop the non-convex constraint ${\text{rank}}\left( {{{\mathbf{C}}_k}} \right) = 1$ in \eqref{sdp problem}. However, based on the SDP rank reduction result in Theorem 3.2 in \cite{ref13}, the optimal solution ${\mathbf{C}}_k^*$ naturally satisfies the rank-1 property. Thus, we can obtain the optimal solution ${\mathbf{c}}_k^*$ for \eqref{approx problem} by SVD on ${\mathbf{C}}_k^*$. Finally, the CPU informs all APs of the precoding vector ${{\mathbf{w}}_k}$ that contains the power information.

Unfortunately, the centralized scheme still involves extensive signaling overhead and high complexity. To this end, an alternative distributed transmission scheme is given in the next subsection, which is easily implementable in practice.

\subsection{Distributed Transmission Scheme}\label{dis_scheme}
To determine the transmission scheme in a distributed approach, we exploit the separable structure of the objective function of \eqref{approx problem}. Moreover, the dual decomposition method is adopted to tackle the coupling constraint.

First, we assume that ${\mathbf{h}}_k^{\text{H}}{{\mathbf{N}}_k}{{\mathbf{c}}_k}$ in \eqref{approx problem} is a real number without loss of generality and neglect the term ‘1’ in log function under high SNR assumption. As a result, the objective function can be transformed as $\max \;\sum\nolimits_k {\ln \left( {{\mathbf{h}}_k^{\text{H}}{{\mathbf{N}}_k}{{\mathbf{c}}_k}} \right)}$, and the corresponding Lagrange function can be expressed as
\begin{eqnarray}
	\begin{aligned}
		\mathcal{L}\left( {{{\mathbf{c}}_1},{{\mathbf{c}}_2}, \ldots {{\mathbf{c}}_K},{\boldsymbol{\lambda }}} \right) = & - \sum\limits_{k = 1}{\ln \left( {{\mathbf{h}}_k^{\text{H}}{{\mathbf{N}}_k}{{\mathbf{c}}_k}} \right)}  + \\ &\sum\limits_{l:\left| {{\mathcal{D}_l}} \right| > 0} {{\lambda _l}\big( {\sum\limits_{k \in {\mathcal{D}_l}} {{{\left\| {{{\mathbf{N}}_{k,l}}{{\mathbf{c}}_k}} \right\|}^2}}  - {\rho _{\max }}} \big)},
	\end{aligned}
\end{eqnarray}
where ${\boldsymbol{\lambda }}$ is the dual variable associated with the per AP power constraint.

The dual decomposition method solves the dual problem
\begin{eqnarray}\label{dual problem}
	\begin{aligned}
		\max                            \;\; & g\left( {\boldsymbol{\lambda }} \right) \triangleq \mathop {\inf }\limits_{{{\mathbf{c}}_1}, \ldots ,{{\mathbf{c}}_K}} \;\mathcal{L}\left( {{{\mathbf{c}}_1}, \ldots {{\mathbf{c}}_K},{\boldsymbol{\lambda }}} \right) \\
		{\text{s}}{\text{.t}}{\text{.}} \;\; & {\boldsymbol{\lambda }}\succeq{\mathbf{0}}                                                                                                                                                                             \\
	\end{aligned}
\end{eqnarray}
via the projected gradient ascent algorithm in an iterative approach. In the $n$-th iteration, the CPU sets $\lambda _l^{\left( n \right)}$ for each serving AP. To derive the gradient $\nabla g( {\boldsymbol{\lambda}^{(n)}} )$, a specific AP in each $\mathcal M_k$ is required to collect necessary CSI between $\mathcal M_k$ and $\mathcal C_k$, and to solve the sub-problem


\begin{equation}\label{subproblem}
	\mathop {\min }\limits_{{{\mathbf{c}}_k}} \;\;\sum\limits_{l \in {\mathcal{M}_k}} {\lambda _l^{\left( n \right)}{{\left\| {{{\mathbf{N}}_{k,l}}{{\mathbf{c}}_k}} \right\|}^2}}  - \ln \left( {{\mathbf{h}}_k^{\text{H}}{{\mathbf{N}}_k}{{\mathbf{c}}_k}} \right)
\end{equation}
in a distributed approach. Since the objective function is convex, and based on the first-order condition, the closed-form optimal solution can be derived as
\begin{equation}\label{subproblem solution}
	{\mathbf{c}}_k^* = \frac{{{{\mathbf{A}}_k}{\mathbf{N}}_k^{\text{H}}{{\mathbf{h}}_k}}}{{\sqrt {{\mathbf{h}}_k^{\text{H}}{{\mathbf{N}}_k}{{\mathbf{A}}_k}{\mathbf{N}}_k^{\text{H}}{{\mathbf{h}}_k}} }},
\end{equation}
where ${{\mathbf{A}}_k}{\text{ = }}{\Big( {\sum\nolimits_{l \in {\mathcal{M}_k}} {2\lambda _l^{\left( n \right)}{\mathbf{N}}_{k,l}^{\text{H}}{{\mathbf{N}}_{k,l}}} } \Big)^{ - 1}}$. By substituting $\mathbf{c}_k^*$ into \eqref{dual problem}, the partial derivative of $g\left( {{{\boldsymbol{\lambda }}^{\left( n \right)}}} \right)$ w.r.t. $\lambda _l^{\left( n \right)}$ can be obtained as
\begin{equation}
	\frac{{\partial g}}{{\partial \lambda _l^{\left( n \right)}}} = \left( {\sum\limits_{k \in {\mathcal{D}_l}} {{{\left\| {{{\mathbf{N}}_{k,l}}{\mathbf{c}}_k^*} \right\|}^2}}  - {\rho _{\max }}} \right).
\end{equation}
The CPU collects $\partial g/\partial \lambda _l^{\left( n \right)}$ sent by the APs and updates ${\boldsymbol{\lambda }}$ with step-size $\alpha $ as
\begin{equation}
	{{\boldsymbol{\lambda }}^{\left( {n + 1} \right)}} = {\left[ {{{\boldsymbol{\lambda }}^{\left( n \right)}} + \alpha \nabla g\left( {{{\boldsymbol{\lambda }}^{\left( n \right)}}} \right)} \right]_ + },
\end{equation}
where ${\left[ {\mathbf{x}} \right]_ + } \triangleq \max \left\{ {{\mathbf{x}},{\mathbf{0}}} \right\}$ element-wise. Since the dual objective function is always concave, ${{\boldsymbol{\lambda }}^{\left( n \right)}}$ is guaranteed to converge to the optimal solution ${{\boldsymbol{\lambda }}^*}$ \cite{ref14}. Moreover, it is obvious that strong duality holds for the optimization problem, and we can derive ${\mathbf{c}}_k^*$ by substituting ${{\boldsymbol{\lambda }}^*}$ into \eqref{subproblem solution}.

In the proposed scheme, only scalars $\lambda _l^{\left( n \right)}$ and $\partial g/\partial \lambda _l^{\left( n \right)}$ are exchanged before downlink data transmission, and the CSI exchange between the CPU and the APs is completely avoided. Besides, since the AP clusters can calculate the precoder in a distributed approach, both data-related overhead and complexity can also be reduced.

\section{Simulation Results}
This section presents the simulation results of the proposed scheme and compares it with existing works. In the simulation, we consider the network area consisting of $L=100$ APs with $N=4$ antennas and $K=20$ users. The APs are deployed on a uniform grid with a minimum spacing of 100m. The system works at 2GHz, and the pathloss is modeled as
\begin{equation}
	\beta_{k,l}[\mathrm{dB}]=-30.5-36.7 \log _{10}\left(\frac{d_{k,l}}{1 \mathrm{m}}\right)+F_{k,l},
\end{equation}
where $d_{k,l}$ is the distance between AP $l$ (taking 10m height difference into account) and user $k$, and $F_{k,l}\sim\mathcal N(0,4^2)$ is the shadow fading. The shadowing terms are correlated as
\begin{equation*}
	\setlength{\nulldelimiterspace}{0pt}
	\mathbb{E}\left\{F_{k,l} F_{i,j}\right\} = \left\{\begin{IEEEeqnarraybox} [\relax] [c] {l's}
		{4^{2} 2^{-\delta_{k,i} / 9 \mathrm{~m}} }, &$l=j, $\\
		{0}, &$l \neq j,$
	\end{IEEEeqnarraybox}\right.
\end{equation*}
where $\delta_{k,i}$ is the distance between user $k$ and user ~$i$. The spatial correlation is generated via the Gaussian local scattering model with $15^{\circ}$ angular standard deviation \cite{ref9}. Moreover, the normalized maximum transmit power and the step-size is set as $\rho_{\mathrm{max}}=94\mathrm{dB}$ and $\alpha=0.05$, respectively. All simulation results are averaged over 1000 channel realizations.

\begin{figure}
	\centering
	\includegraphics[width=2.2in]{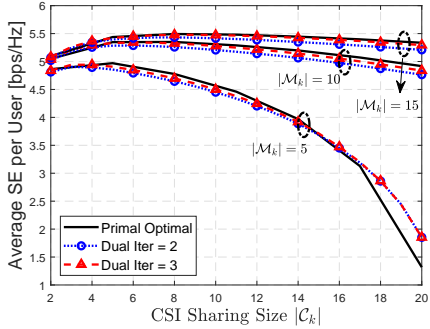}
	\caption{SE performance under different system parameters.}
	\label{fig2}
\end{figure}

Fig. \ref{fig2} demonstrates the performance of the proposed distribution scheme under different system parameters. First, it can be observed that a higher SE can be achieved through the enlargement of AP clusters. However, the performance gain increases slowly when the $|\mathcal M_{k}|$ is large enough. Moreover, the SE increases slowly at first and then decreases with the size of CSI sharing sets $|\mathcal C_k|$, and this trend is obvious when the spatial freedom is insufficient, i.e., $|\mathcal M_{k}|=5$. We can infer that the elimination of all users' interference is not necessary in a user-centric CF network, and the proposed scheme works well when the AP cluster is small. Finally, we verify the convergence property of the distributed scheme, and the primal problem is optimally solved as a comparison\footnote{Since the objective function of \eqref{approx problem} is just an approximation of the actual SE, the dual method may achieve a higher SE in special cases.}. The results show that the proposed scheme based on the dual gradient method can achieve a better performance with only 2-3 iterations, which significantly reduces the interaction between the CPU and the APs.

\begin{figure}
	\centering
	\includegraphics[width=2.2in]{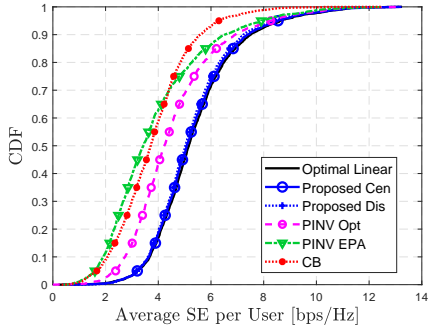}
	\caption{SE performance under different transmission schemes.}
	\label{fig3}
\end{figure}

Fig. \ref{fig3} compares the performance of different transmission schemes. The simulation parameters are set as $|\mathcal C_k|=5$, $|\mathcal M_{k}|=10$, and the distributed scheme performs 2 iterations. As a comparison, we consider the decoupling schemes that select the precoder from the pseudo-inverse  matrix (\emph {'PINV'}), followed by an equal or optimal power allocation strategy (\emph {'EPA'} or \emph{'Opt'}). Besides, the algorithm in \cite{ref16} is utilized to obtain the near-optimal SE performance under the linear precoder. The results show that our scheme achieves a higher SE than the comparison schemes that adopt a decoupling method, which reflects the advantage of the joint design. Finally, we can observe that the distributed scheme achieves the same performance as the centralized one, which is closed to the near-optimal one.

Finally, we compare the date-related overhead of the distributed scheme with the centralized one, which mainly focuses on the link between the CPU and a single AP. For the distributed scheme, the total overhead is ${\tau _{\text{d}}}\bar KB$ bits \cite{ref1}, where $\bar K$ denotes the average number of users served by each AP, and $B$ represents the number of bits per symbol. For the centralized scheme, the CPU calculates the  $N$-dimensional transmitted signals, which is quantized with  $A$ bits for I/Q channels, respectively, and the total overhead is $2{\tau _{\text{d}}}NA$ bits. In the user-centric cell-free network, we have $\bar K \ll K$. Consider the simulation parameters above and $A = 8$, $B = 4$, our distributed scheme reduces the signaling overhead by 87.5\%.

\section{Conclusion}
This letter studies the downlink transmission scheme for the user-centric CF system. To avoid extensive information exchange between the CPU and the APs, we propose a distributed solution that jointly optimizes the precoding and power. With 2-3 iterations, the proposed scheme performs close to the centralized one and has a better performance than the schemes that adopt a decoupling method. Meanwhile, it significantly reduces the signaling overhead and complexity.

\ifCLASSOPTIONcaptionsoff
	\newpage
\fi

\bibliographystyle{IEEEtran}
\bibliography{IEEEabrv,mybib}

\end{document}